\documentclass{PoS}

\usepackage{amsmath}
\usepackage{braket}
\usepackage{xcolor}
\usepackage{lineno}

\title{Equation of state in (2+1) flavor QCD at high temperatures}

\ShortTitle{Equation of state in (2+1) flavor QCD at high temperatures}

\author{\speaker{Johannes Heinrich Weber}
\\
        Michigan State University\\
        Department of Computational Mathematics, Science and Engineering\\
        E-mail: \email{weberjo8@msu.edu}}

\author{Alexei Bazavov\\
       Michigan State University\\
       Department of Computational Mathematics, Science and Engineering\\
       Department of Physics and Astronomy\\
       E-mail: \email{bazavov@msu.edu}}

\author{Peter Petreczky\\
       Brookhaven National Laboratory\\
       Department of Physics\\
       E-mail: \email{petreczk@quark.phy.bnl.gov}}

\definecolor{red}      {rgb}{0.8,0.0,0.0}
\definecolor{green}    {rgb}{0.0,0.6,0.0}
\definecolor{darkblue} {rgb}{0.0,0.1,0.7}
\definecolor{brown}    {rgb}{0.6,0.1,0.0}
\definecolor{gray}     {rgb}{0.6,0.6,0.6}
\definecolor{darkgreen}{rgb}{0.0, 0.545098, 0.0}
\definecolor{orange}   {RGB}{238,80,25}
\definecolor{purple}   {rgb}{0.5,0.0,0.5}
\definecolor{babypink} {rgb}{0.64, 0.44, 0.44}

\definecolor{spartandarkgreen} {RGB}{24, 69, 59}
\definecolor{spartanlightgreen} {RGB}{13, 177, 75}
\definecolor{spartandarkgray} {RGB}{155, 162, 162}
\definecolor{spartanlightgray} {RGB}{153, 80, 84}
\definecolor{mytumgray}{HTML}{B0B0B0}
\definecolor{mytumblue}{HTML}{3071B5}
\definecolor{tumblue}{RGB}{0,101,189}

\newcommand{\al}[1]{\vskip-3ex\begin{align}#1\end{align}}

\newcommand{\bmat}{\left(\begin{array}}
\newcommand{\emat}{\end{array}\right)}

\newcommand{\as}{\alpha_\mathit{s}}

\abstract{
We calculate the Equation of State at high temperatures in (2+1) flavor QCD 
using the highly improved staggered quark (HISQ) action. 
We study the lattice spacing dependence of the pressure at high temperatures
using lattices with temporal extent $N_{\tau}=6,~8,~10$ and $12$ and perform
continuum extrapolations. 
We also give a continuum estimate for the Equation of State up to temperatures 
$T=2$\,GeV, which are then compared with results of the weak-coupling 
calculations. 
We find reasonable agreement with the HTL calculation at 
the highest temperatures and a tension with the EQCD calculation
of a few percent, with the lattice results in the middle between both.
}

\FullConference{XIII Quark Confinement and the Hadron Spectrum - Confinement2018\\
		31 July - 6 August 2018\\
		Maynooth University, Ireland}

\begin{document}

\section{Introduction}

Ab initio calculations of the QCD Equation of State (EoS) at vanishing baryon 
density have been in the focus of research for more than a decade, 
see~\cite{Cheng:2007jq, Bazavov:2009zn, Borsanyi:2010cj, 
Borsanyi:2013bia, Bazavov:2014pvz, Borsanyi:2016ksw, 
Bazavov:2017dsy} and references therein. 
These calculations were performed with different varieties of improved staggered fermions, nearly physical quark masses and 2+1 or 2+1+1 flavors of sea quarks. 
As a result, the EoS in the continuum limit has been determined for 2+1 flavors 
using two different staggered discretizations (HISQ~\cite{Follana:2006rc} and stout) 
up to temperatures of about \(T = 400\,{\rm MeV}\)~\cite{Borsanyi:2013bia, 
Bazavov:2014pvz} and for 2+1+1 flavors using only the stout formulation up to 
temperatures of about \(T = 1\,{\rm GeV}\)~\cite{Borsanyi:2016ksw}.  

Overall the different 2+1 flavor calculations are in good agreement, although there 
is a slight tension at the highest temperatures. 
By comparing the 2+1 and 2+1+1 flavor calculations using the stout formulation it 
was concluded that the contribution from the charm sea is already significant at 
\(T \approx 400\,{\rm MeV}\). 
However, as the presence of the charm quark in the sea may lead to a suppression of 
the other contributions to the EoS, the overall effect may be subject to subtle 
compensations. 
Moreover, the 2+1 and 2+1+1 flavor results are compared in the region where the 
tension between both formulations has been found. 
From this discussion it is clear that dynamical charm quarks should be included in 
the lattice determination of the EoS at higher temperatures for use in phenomenology 
and that an independent calculation is desirable. 
Efforts in this direction with the HISQ formulation are still on-going~\cite{Bazavov:2013pra}. 
In particular, the large discretization errors associated with the charm quark mass 
strongly suggest that such a calculation should be performed with the HISQ action. 
Yet the 2+1 flavor EoS is needed with higher precision for a robust, quantitative 
determination of the correct charm contribution to the pressure including all of 
the subtle effects. 

For a comparison to weak-coupling calculations a lattice determination 
without the charm quark may actually be advantageous. 
This is evident from the absence of massive quark degrees of freedom in the 
weak-coupling calculations~\cite{Laine:2006cp, Haque:2014rua} at the highest 
available order, 
NNNLO\footnote{The effects of finite charm quark mass are known only up to NLO.}. 
At temperatures in the range between \(T \approx 400\,{\rm MeV}\) and some multiple 
of the charm quark mass, \(m_c\equiv m_c(m_c)=1.277(10)\,{\rm GeV}\)~\cite{PPJHW:as2018}, corrections  
due to the finite charm quark mass are almost certainly relevant and large. 
Unless the applicability of weak coupling is firmly established 
for the respective observables at these temperatures, there is a risk of 
conflating the uncertainties associated with the temperature scale with those of 
the charm mass corrections. 
Precise results with 2+1 flavors at high temperatures from the lattice alleviate 
the concerns regarding applicability of weak coupling considerably and establish 
a robust baseline for determining the actual charm quark contribution. 

For these reasons we extend the previous calculations of the 2+1 flavor EoS using 
the HISQ formulation to higher temperatures. 
We use lattices with larger light sea quark mass, \(m_l=m_s/5\) (compared to 
\(m_l=m_s/20\) in~\cite{Bazavov:2014pvz}), and discuss the light quark mass 
dependence of the trace anomaly.   
We discuss the cutoff dependence of the trace anomaly and the pressure in detail 
and determine the continuum limit or continuum estimates using different strategies. 
The rest of this paper is organized as follows. 
In \mbox{Sec.}~\ref{sec:lattice}, we discuss the details of the lattice calculations. 
In \mbox{Sec.}~\ref{sec:traceanomaly}, we discuss the trace anomaly and its cutoff effects. 
In \mbox{Sec.}~\ref{sec:pressure}, we obtain results for the pressure and scrutinize its cutoff effects. 
We compare to weak-coupling results in \mbox{Sec.}~\ref{sec:weak} and conclude in \mbox{Sec.}~\ref{sec:conclusions}. 

\section{Details of the lattice calculations}\label{sec:lattice}

The objective of our research is extending the calculation of the 2+1 flavor QCD 
Equation of State in~\cite{Bazavov:2014pvz} to higher temperatures and compare the 
EoS to results obtained in the weak-coupling limit. 
Following~\cite{Bazavov:2014pvz} we use the tree-level improved Symanzik gauge action and the Highly Improved Staggered Quark (HISQ) action for quarks. 
We use gauge ensembles at finite temperature with \(N_\tau=12,\ 10,\ 8,\ 6\) and 
\(4\), which have been generated for studies of the pseudocritical 
temperature~\cite{Bazavov:2011nk} and the EoS~\cite{Bazavov:2014pvz} by 
the HotQCD collaboration, and for studies of the entropy shift due to a heavy quark 
in the thermal medium~\cite{Bazavov:2016uvm} and 
color screening~\cite{Bazavov:2018wmo} by the TUMQCD collaboration. 

\begin{figure*}
\includegraphics[width=7.5cm]{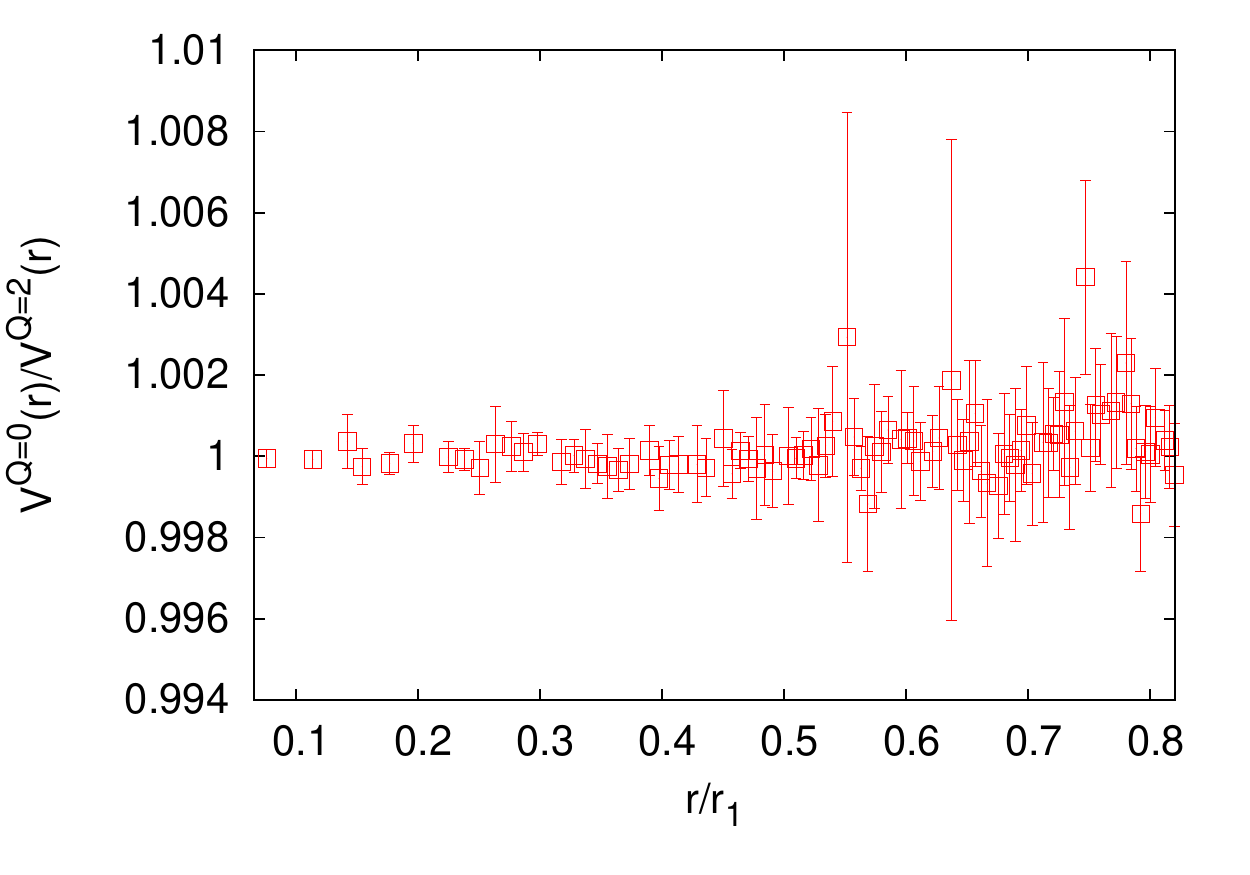}
\includegraphics[width=7.5cm]{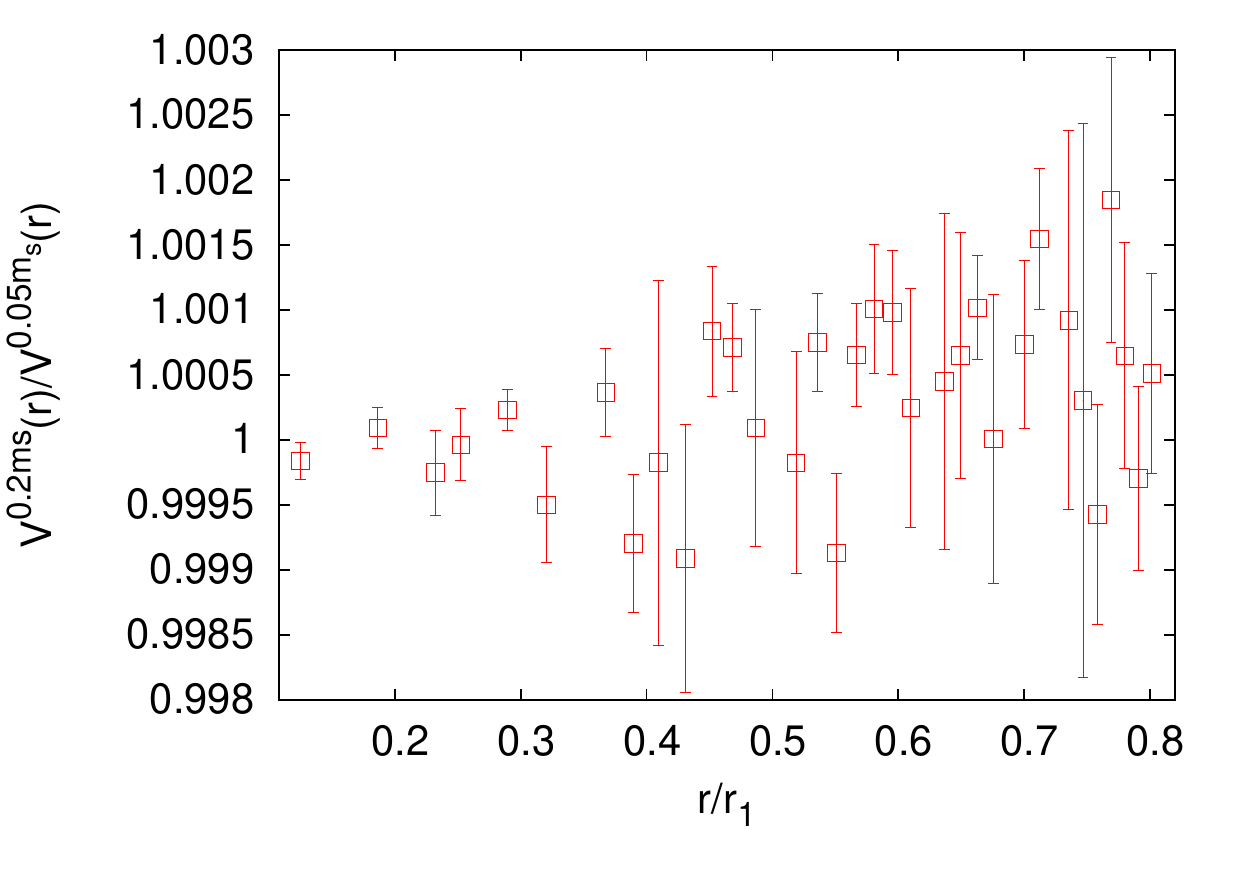}
\caption{
The static energy for different topological charges \(Q=0\) or \(2\) with 
\(\beta=8.4\) (left) or light quark masses \(m_s/5\) or \(m_s/20\) with 
\(\beta=7.825\) (right) is consistent at 0.1\% level for \(r \leq 0.8r_1\). 
We see a consistent trend towards larger values of the static energy for the larger mass, but no trend for variation of \(Q\).
}
\label{fig:pot}
\end{figure*}

In order to cancel the UV divergences in the trace anomaly at \(T>0\), 
we subtract the trace anomaly at \(T=0\) using the same bare parameters. 
We use gauge ensembles at \(T=0\) from the HotQCD 
collaboration~\cite{Bazavov:2014pvz, Bazavov:2011nk} with a pion mass of 
\(m_\pi = 160\,{\rm MeV}\) in the continuum limit. 
Since the corresponding lattice spacings are insufficient for reaching higher 
temperatures with small discretization errors, \mbox{i.e.} large \(N_\tau\), 
we generated new ensembles with finer lattice spacing but larger pion mass, 
\(m_\pi = 320\,{\rm MeV}\)~\cite{Bazavov:2017dsy}. 
We use the rational hybrid Monte Carlo at five values of the coupling, 
\(\beta=10/g^2\), see \mbox{Tab.}~\ref{tab:tzero} for the parameters. 
\begin{table}\center
\begin{tabular}{|c|c|c|c|c|}
\hline
$\beta$ &  $m_s$  &  vol    &  a [fm]  &  \# traj.  \\
\hline
7.030   & 0.03560 & $48^4$  &  0.08253 &  1890 \\
7.825   & 0.01542 & $64^4$  &  0.04036 &  1265 \\
8.000   & 0.01299 & $64^4$  &  0.03469 &  3927 \\
8.200   & 0.01071 & $64^4$  &  0.02924 &  3927 \\
8.400   & 0.00887 & $64^4$  &  0.02467 &  3927 \\
\hline
\end{tabular}
\caption{The parameters of the zero temperature gauge ensembles.}
\label{tab:tzero} 
\end{table}
For the three finest ensembles, \(\beta\geq8\), topological tunneling only takes 
place during pre-thermalized molecular dynamics evolution, but is too 
suppressed after thermalization.
We have generated multiple streams of MD trajectories with the same lattice parameters  
that randomly stall in different topological sectors, \(Q \in \{0,1,2\}\). 
We find among the observables contributing to the EoS only for the light quark 
condensate \(\braket{\bar\psi \psi}_l\) a statistically significant dependence 
on the topological charge \(Q\). 
The respective changes of the light quark contribution to the trace 
anomaly are smaller than the statistical error of the much larger 
contribution from the gauge action. 
We show in the left panel of \mbox{Fig.}~\ref{fig:pot} that the static energy 
is consistent for different values of \(Q\) in the finest ensemble \(\beta=8.4\). 
For the other ensembles the picture is similar.  
We conclude that systematic errors due to frozen topology are numerically 
irrelevant for our study. 

We fix the lattice scale with the static energy, \mbox{i.e.} we calculate \(r_1\) and \(r_2\) defined via 
\al{\label{eq:scale}
&r^2 \left . \frac{d V(r)}{d r} \right|_{r=r_1}=1,& 
&r^2 \left .\frac{d V}{d r}\right |_{r=r_2}=\frac{1}{2},
}
where \(r_1=0.3106\,{\rm fm}\) and \(r_2=0.1413\,{\rm fm}\).
For the two ensembles corresponding to coarser lattices we can directly compare to
 previous calculations by HotQCD with a smaller pion mass. 
This permits us to determine quark mass effects in the lattice scale as well as in 
the thermodynamic observables. 
The static energy changes slightly for the larger quark mass, see \mbox{Fig.}~\ref{fig:pot}, 
and thus, the lattice scale \(r_1\) is 0.7\% or 1.6\% lower \(\beta=7.03\) or \(7.825\), 
respectively. 
Since the scale \(r_2\) is only 0.3\% lower for \(\beta=7.825\) we use \(r_2\) to 
set the scale for the finer ensembles, which is consistent with the non-perturbative 
beta function of~\cite{Bazavov:2014pvz}. 
The observed variation of the static energy due to the frozen topology charge is 
smaller than the quark mass dependence and safely covered within the statistical errors. 
Cominbing the new and old results for the lattice scale we obtain a consistent beta 
function with smaller uncertainty~\cite{Bazavov:2017dsy}, which we parameterize with 
an Allton-type Ansatz 
\al{
&\frac{r_1}{a} =\frac{c_0 f(\beta) + c_2 (10/\beta) f^3(\beta)}{1+d_2(10/\beta) f^2(\beta)},&
& f((\beta) = \left(\frac{10b_0}{\beta}\right)^{-\frac{b_1}{2b_0^2}} e^{-\frac{\beta}{20b_0}},
}
where \(b_0=9/(16\pi^2)\) and \(b_1=1/(4\pi^4)\). 
The other coefficients \(c_0=43.12(18)\), \(c_2=347008(32131)\) and 
\(d_2=5584(599)\) are determined with a fit, see~\cite{Bazavov:2017dsy}. 
In our analysis we combine ensembles with different quark masses, 
using the ensembles with larger pion mass for the UV subtraction only 
at high temperatures, \(T \gtrsim 470\,{\rm MeV}\), where numerical 
effects of \(m_l\) are known to be small for \(m_l<0.4m_s\)~\cite{Borsanyi:2010cj}.
Due to \(T=1/(aN_\tau)\) these ensembles provide access up to \(T = 2\,{\rm GeV}\) 
with \(N_\tau=4\).

\section{Trace anomaly}\label{sec:traceanomaly}

\begin{figure*}
\includegraphics[width=7.5cm]{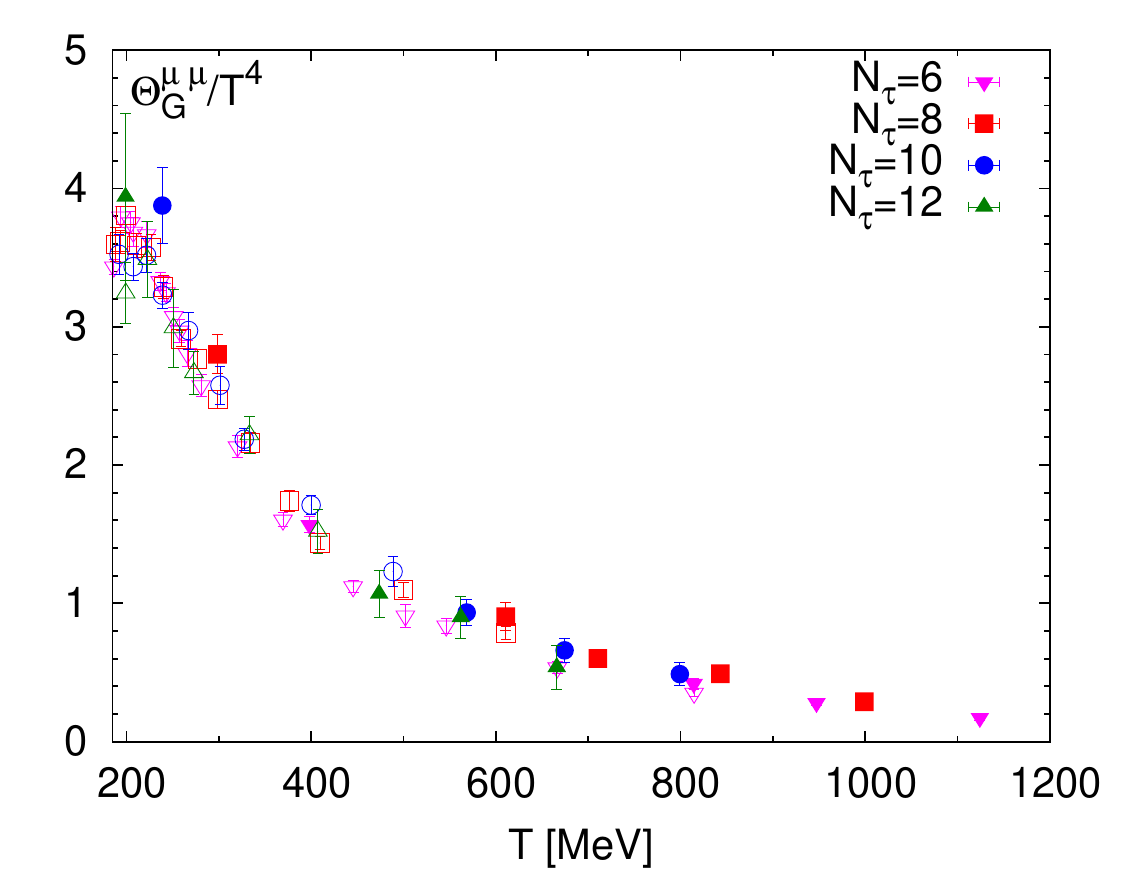}
\includegraphics[width=7.5cm]{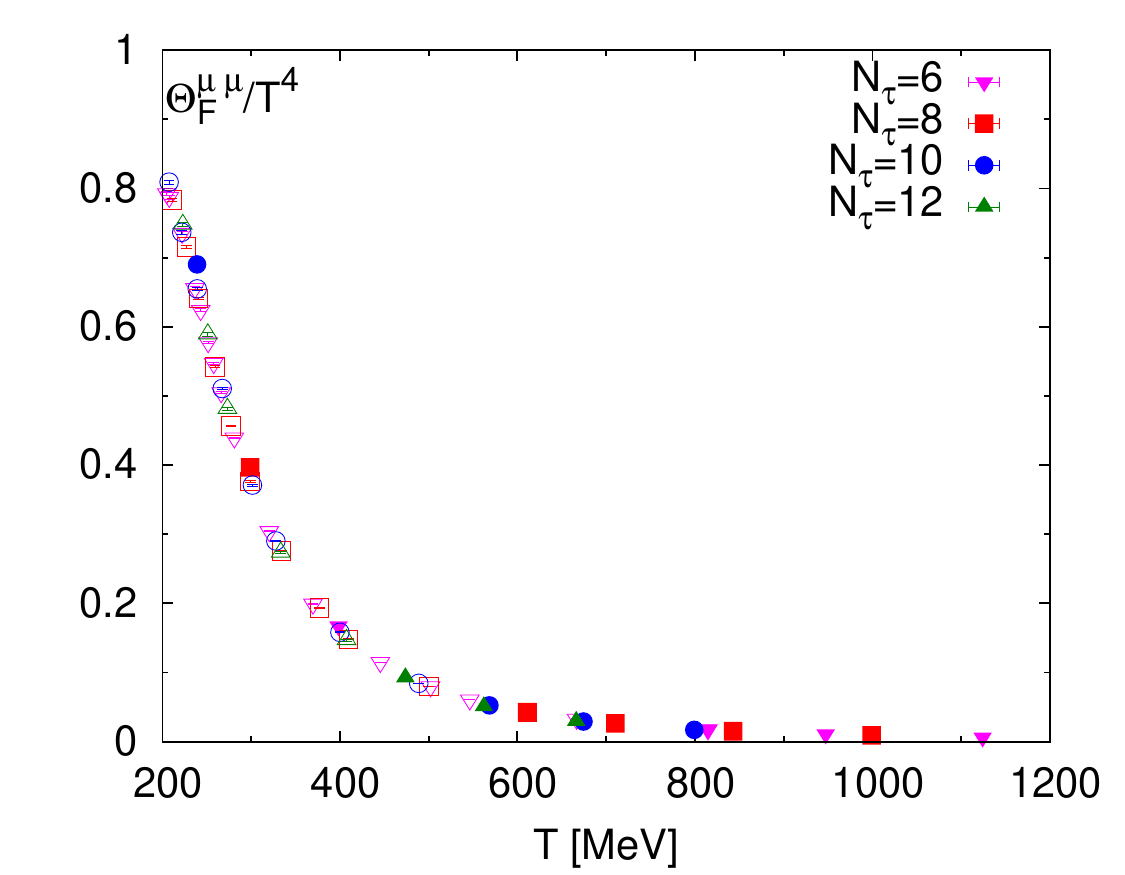}
\caption{The gauge part (left) and the fermion part (right) of the trace anomaly obtained with HISQ action.
The open symbols correspond to HotQCD results with 
$m_l/m_s=1/20$~\cite{Bazavov:2014pvz}, while the filled symbols correspond to 
$m_l/m_s=1/5$.
}
\label{fig:e3p_HISQ}
\end{figure*}

In order to obtain the 2+1 flavor QCD Equation of State we first calculate the 
trace anomaly, \mbox{i.e.} the trace of the energy-momentum tensor, 
\(\Theta^{\mu\mu} =\epsilon-3p\) on the lattice. 
\(\Theta^{\mu\mu}\) can be expressed in terms of the expectation values of the gauge 
action and the scalar condensates of the quark flavors in the sea, \mbox{i.e.} light 
and strange. 
Since these densities diverge linearly in the continuum limit, we consider differences between the results at zero temperature and at finite temperature. 
Finally, we account for multiplicative renormalization bu rescaling these differences with the non-perturbative beta function \(R_\beta\) and mass renormalization function \(R_m\). 
For the HISQ action we may write 
\al{
 \Theta^{\mu\mu}(T)
 &=\Theta^{\mu\mu}_G(T)+\Theta^{\mu\mu}_F(T), \label{eq:Tmumu}
 \\
 \frac{\Theta^{\mu\mu}_G(T)}{T^4},
 &=R_\beta\left\{\braket{S_G}_{T=0}-\braket{S_G}_{T>0}\right\} N_\tau^4,
 \label{eq:TGmumu}\\
 \frac{\Theta^{\mu\mu}_F(T)}{T^4}
 &=-R_\beta R_m\left\{
 2m_l\left(\braket{\bar\psi\psi}_{l,T=0}-\braket{\bar\psi\psi}_{l,T>0}\right)
 +m_s\left(\braket{\bar\psi\psi}_{s,T=0}-\braket{\bar\psi\psi}_{s,T>0}\right)
 \right\} N_\tau^4
 \label{eq:TFmumu}
}
using the same notation as in~\cite{Bazavov:2014pvz}. 
\(R_\beta\) and \(R_m\) are defined via 
\al{
&R_\beta(\beta)=\frac{r_1}{a} \left(\frac{d (r_1/a)}{d \beta}\right)^{-1},
&&
R_m(\beta)=\frac{1}{m_s(\beta)} \left(\frac{d m_s(\beta)}{d \beta}\right),
}
\(\Theta^{\mu\mu}_G\) is much larger than \(\Theta^{\mu\mu}_F\). 
We show the two contributions to the trace anaomaly in \mbox{Fig.}~\ref{fig:e3p_HISQ}. 
The gauge field contribution, which we show in the left panel, is systematically larger for the ensembles with larger sea quark mass. 
Although we cannot correct for this implicit quark mass dependence, the difference 
is covered by the statistical errors for temperatures above 
\(T \gtrsim 400\,{\rm MeV}\), \mbox{i.e.} in the range where we use these data. 
For \(N_\tau \geq 8\) discretization errors are quite visible, but mostly covered by 
the rather large statistical errors in this temperature window.
The fermion contribution, which we show in the right panel, has an explicit dependence on the sea quark mass. 
As we use a value \(m_l=m_s/20\) instead of \(m_l=m_s/5\) in \eqref{eq:TFmumu} 
to account for this, we cannot resolve a residual quark mass dependence in 
\(\Theta^{\mu\mu}_F\) anymore for \(T \gtrsim 400\,{\rm MeV}\). 
Moreover, statistical errors and cutoff effects are very small. 
We separately discuss the trace anomaly for low and high temperatures. 

\begin{figure*}
\includegraphics[width=7.5cm]{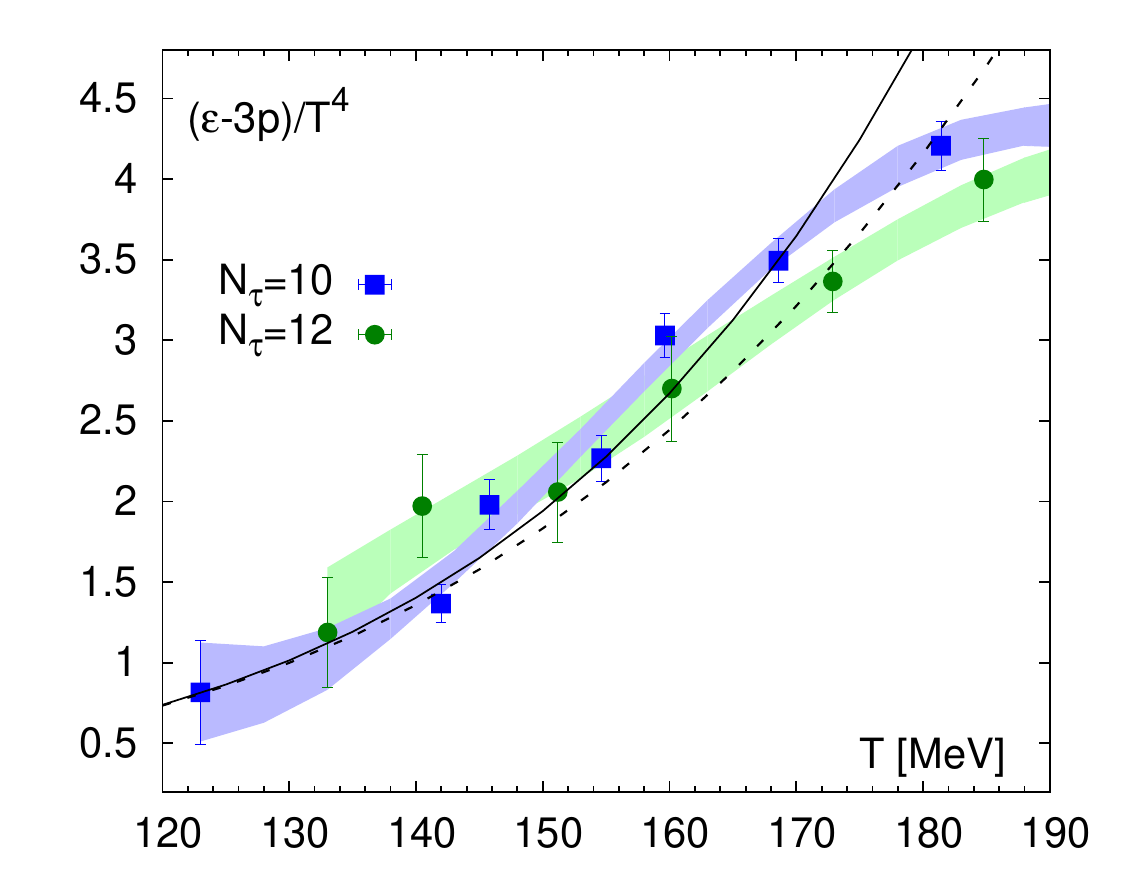}
\includegraphics[width=7.5cm]{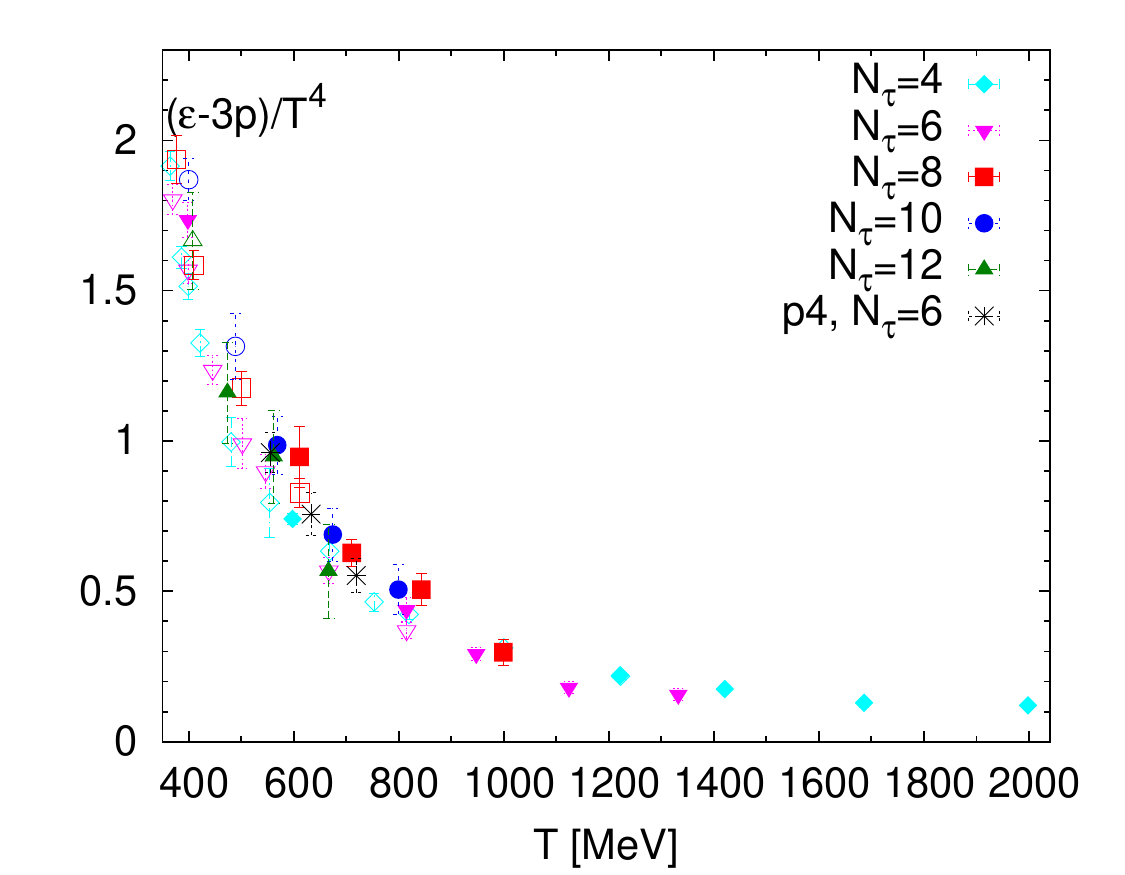}
\caption{Left: 
The trace anomaly calculated in the low temperature region. 
The bands correspond to interpolations. 
The dashed line corresponds to HRG-PDG, while the solid line corresponds to 
HRG-QM.
Right: 
The numerical results for the trace anomaly in the high temperature region 
at two different quark masses.
The bursts correspond to calculations with p4 action and 
$m_l=m_s/10$~\cite{Cheng:2007jq}.
}
\label{fig:e3p_lowhigh}
\end{figure*}

We revisit the low temperature results for the trace anomaly since we require 
higher precision at low temperatures in order to take the continuum limit in terms 
of the pressure. 
In this study we use gauge ensembles at low temperatures with high statistics that have been generated for a study of the Polyakov loop in the crossover region by the TUMQCD collaboration~\cite{Bazavov:2016uvm}. 
These ensembles provide access to \(T = 123\,{\rm MeV}\) with \(N_\tau=10\) and to 
\(T = 133\,{\rm MeV}\) as well as \(T = 140\,{\rm MeV}\) with \(N_\tau=12\). 
We show our low temperature results for the trace anomaly with \(N_\tau=12\) and \(10\) in the left panel of \mbox{Fig.}~\ref{fig:e3p_lowhigh} together with bands obtained from interpolating splines whose errors have been estimated from the bootstrap method. 
The difference between both in the window \(160\,{\rm MeV} < T < 180\,{\rm MeV}\) is 
indicative of rather large cutoff effects.
We compare our results directly to predictions of hadron resonance gas (HRG) models. 
Namely, we consider a model with only the states listed by the PDG -- labeled 
HRG-PDG -- and a model including also missing states, \mbox{i.e.} states predicted 
by the quark model that have not been confirmed experimentally -- labeled HRG-QM. 
We find good agreement of both HRG models with the data up to 
\(T < 140\,{\rm MeV}\). 
For higher temperatures the difference between both models is significantly and 
they cannot describe the data well. 

At high temperatures \(T > 400\,{\rm MeV}\), discretization errors and quark mass dependendence are very much suppressed. 
We show the HISQ results together with older results obtained with the p4 action 
and a quark mass of \(m_l=m_s/10\). 
Whereas the results with \(N_\tau=4\) and \(N_\tau=6\) are systematically lower, 
the results for \(N_\tau\geq8\) do not depend on \(N_\tau\) or the light quark 
mass within their uncertainties. 
The smallness of the cutoff effects can be understood from the weak-coupling 
picture, where the trace anomaly has discretization errors starting at three-loop order, \mbox{i.e.} \(\mathcal{O}(\as^2)\). 
Since the running of \(\as\) is controlled by the temperature, discretization 
errors are strongly suppressed at high temperatures and their respective 
temperature dependence is mild, \mbox{i.e.} logarithmic. 

\section{Pressure}\label{sec:pressure}

\begin{figure*}
\includegraphics[width=7.5cm]{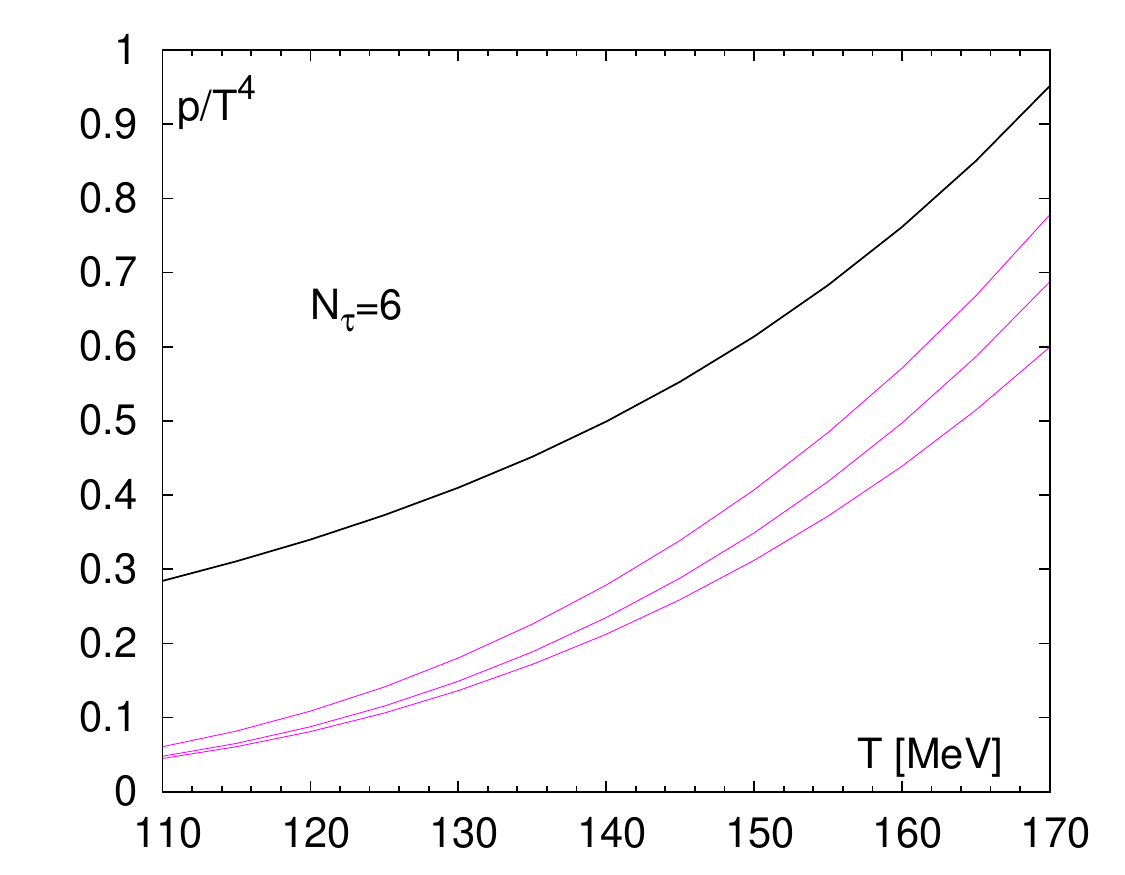}
\includegraphics[width=7.5cm]{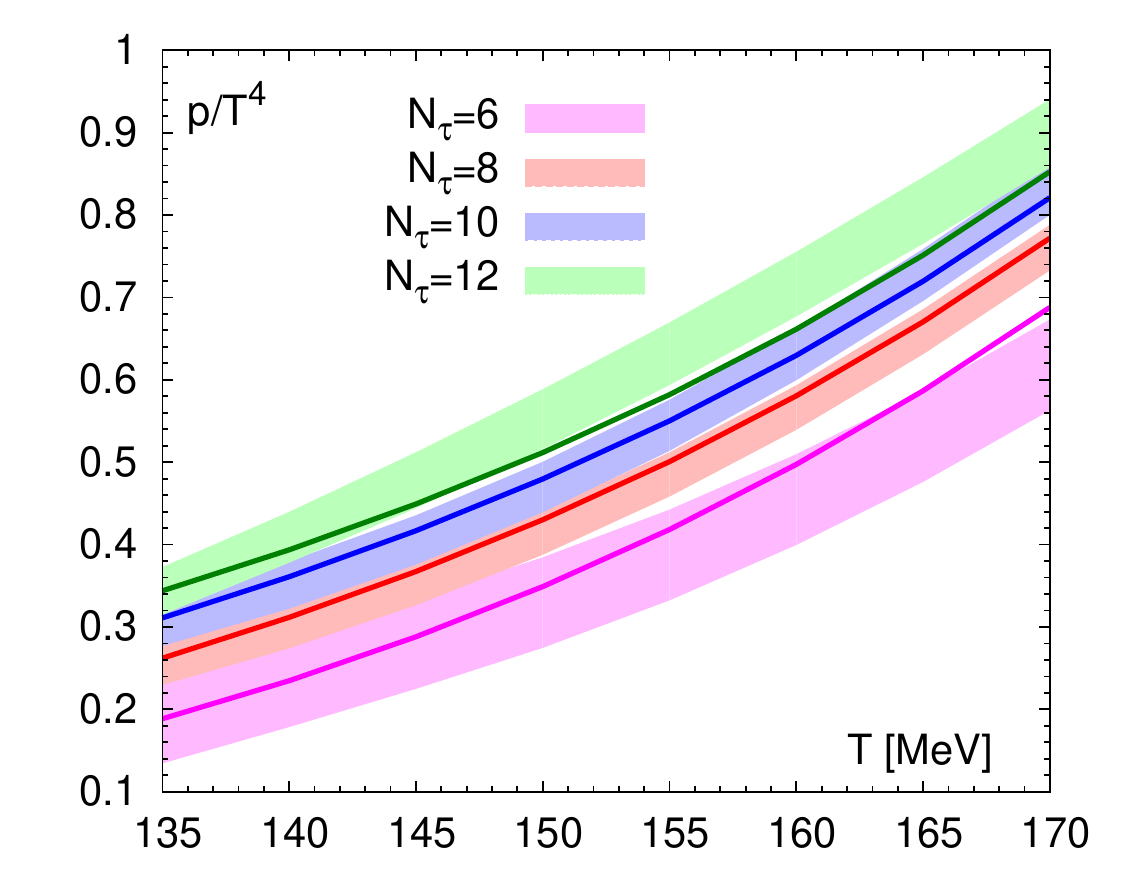}
\caption{Left:
The pressure with \(N_\tau=6\) for HRG with different distortions of the hadron spectrum.
Right:
The lines correspond to HRG with distorted hadron spectrum (see text).
}
\label{fig:p_low}
\end{figure*}

Eventually we obtain the Equation of State and the pressure from \(\Theta^{\mu\mu}\) via the integral method, 
\al{\label{eq:integral.method}
 \frac{p}{T^4} - \frac{p_0}{T_0^4} = \int\limits_{T_0}^T dT^\prime \frac{\Theta^{\mu\mu}}{T^{\prime\,5}}.
}
Here, we assume that we know the pressure \(p(T_0)\) at the reference temperature 
\(T_0\). 
If \(T_0\) is sufficiently small, the pressure may be set to zero or taken from a 
calculation using an HRG model. 
For the temperature range in our simulations, we have to adhere to the latter case. 
Moreover, we have to account in the HRG for the cutoff effects due to distortion 
of the hadron spectrum with the HISQ formulation. 
In short, for each species of hadrons there are multiple tastes of the same 
hadron with masses that are enlarged by discretization errors to varying degree. 
This distortion is most severe in the sector of pseudo-Goldstone bosons. 
Hadrons with larger masses contribute less to the pressure. 
The effect of this distortion is suppressed in the trace anomaly since hadrons 
with larger masses contribute more strongly to the trace anomaly and partly 
compensate for the distortion. 
We use the HRG-QM where only the pseudo-Goldstone bosons are modified, and 
estimate a systematic uncertainty from the difference to the HRG-QM where all 
ground states or all ground and excited states are modified, 
see~\cite{Bazavov:2017dsy} for a more detailed discussion.
In the left panel of \mbox{Fig.}~\ref{fig:p_low} we show the lattice result for 
the pressure at low temperatures together with a prediction of the HRG with a 
distorted spectrum. 
In the right panel, we show the lattice QCD results obtained via the integral 
method together with the HRG prediction. 
We summarize the reference temperatures and pressures for different \(N_\tau\) 
in \mbox{Tab.}~\ref{tab:pzero}.  
\begin{table}\center
\begin{tabular}{|c|c|c|}
\hline
$N_{\tau}$ & $T_0$ [MeV] & $p(T_0,N_{\tau})$ \\
\hline
6          & 135         & 0.189(54)         \\
8          & 120         & 0.145(22)         \\
10         & 125         & 0.226(23)         \\
12         & 135         & 0.344(29)         \\
\hline
\end{tabular}
\caption{The values of $T_0$ and $p(T_0)$ used to calculate
the pressure for different $N_{\tau}$ (see text).
}
\label{tab:pzero} 
\end{table}
Clearly the dominant cutoff effects in the pressure at low temperatures are due to the taste-symmetry violation in the sector of 
pseudo-Goldstone bosons, \mbox{i.e.} they are of the form \(\mathcal{O}(\as a^2)\) 
for the HISQ action. 
At fixed temperature \(T=1/(aN_\tau)\), we may neglect the running of \(\as\) with 
\(T\) and parameterize cutoff effects by powers of \(1/N_\tau^2\).

\begin{figure*}
\includegraphics[width=7.5cm]{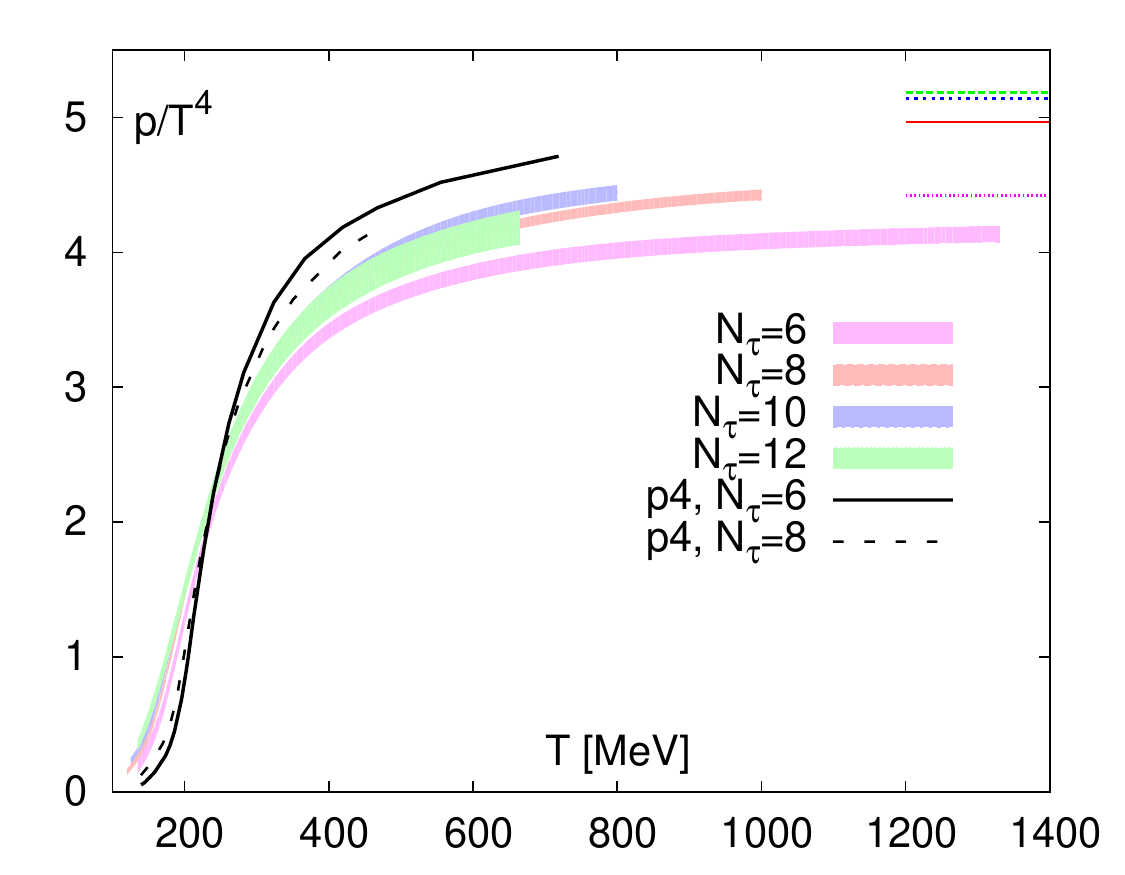}
\includegraphics[width=8cm]{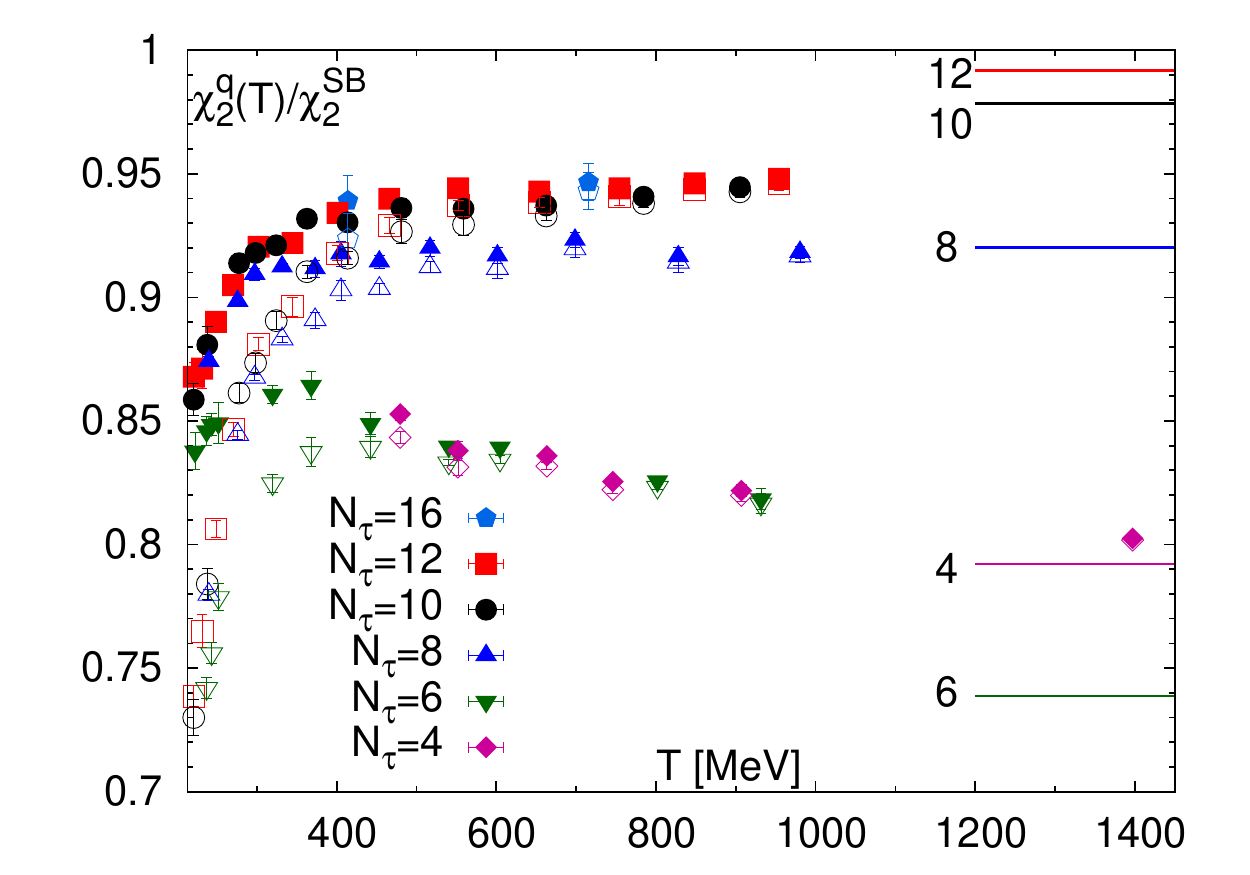}
\caption{Left:
The pressure in the entire temperature range.
The horizontal lines correspond to the free theory result.
Also shown are the results for the pressure obtained with p4 action and
$N_{\tau}=6$ or $8$ \cite{Cheng:2007jq, Bazavov:2009zn}.
Right:
The discretization errors of the second order QNS follow the same pattern. 
\cite{Bazavov:2013uja}.
}
\label{fig:p_chi}
\end{figure*}

For high temperatures the picture is rather different and considerations 
assuming weak coupling apply. 
For the HISQ or p4 actions the pressure and the quark number 
susceptibilities (QNS),
\al{
 &\chi_{2n}^q(T)=\left.\frac{\partial^{2n} p(T,\mu_q)}{\partial \mu_q^2} \right|_{\mu_q=0},
 &&
 n=1,2,\quad q=l,s,
} 
are quite close to the result for the ideal gas limit. 
The dominant discretization errors in the ideal gas limit are starting with a 
one-loop contribution, \mbox{i.e.} \(\mathcal{O}(\as^0)\), and can be parameterized 
as starting with the power \(1/N_\tau^4=(aT)^4\) at fixed temperature. 
Inspection of the lattice results for the pressure and second order QNS shown in \mbox{Fig.}~\ref{fig:p_chi} indicates that the lattice 
results for both are about 15\% below the ideal gas limit for 
\(T \gtrsim 400\,{\rm MeV}\) and follow the same pattern of cutoff effects.  

These considerations open up four different approaches to obtain a continuum 
result for the pressure. 
First, we may follow~\cite{Bazavov:2014pvz} and obtain the continuum limit of the 
trace anomaly and then obtain the continuum pressure via the integral method. 
Second, we may use the integral method at finite \(N_\tau\) to obtain the 
pressure and extrapolate the pressure at fixed temperature to the continuum limit 
using the appropriate parameterizations of the cutoff effects as discussed in 
the preceding paragraphs. 
For an intermediate temperature window, \(200\,{\rm MeV} < T < 400\,{\rm MeV}\), 
the picture is less clear and cutoff effects can be described by \(1/N_\tau^2\), 
\(1/N_\tau^4\) or combinations thereof. 
Continuum results for either parameterization are consistent within errors in this 
window, although the continuum limit for \(1/N_\tau^2\) form is slightly higher 
and the continuum limit for the fits with combined form has large uncertainties 
due to having only one degree of freedom. 
We consider the differences as a measure of the systematic uncertainty and use 
the continuum result for \(1/N_\tau^4\) form as our central value. 
We double the corresponding statistical error in this window, since this is 
numerically even more conservative than adding the systematic error in quadrature. 
For temperatures above \(T > 670\,{\rm MeV}\) only data with \(N_\tau \leq 10\) is 
available and we may consistently extrapolate with controlled uncertainties using 
the same form. 
Above \(T>800\,{\rm MeV}\) we have only \(N_\tau \leq 8\) and cannot control 
the uncertainty of an extrapolation, but provide only a continuum estimate. 
For \(T \approx 800\,{\rm MeV}\) a tentative extrapolation using only 
\(N_\tau \leq 8\) yields within errors the same coefficients as an extrapolation using 
\(N_\tau \leq 10\). 
Thus, the continuum estimate seems to be on rather solid ground. 
This is not completely surprising, since we expect that the coefficients have 
only a mild temperature dependence due to the logarithmic running of the coupling. 
In the same spirit, we use the coefficient of \(1/N_\tau^4\) at \(T = 1\,{\rm GeV}\) 
to estimate the correction for the pressure with \(N_\tau=6\) at higher temperatures, 
up to \(T = 1.33\,{\rm GeV}\). 
Third, we may use our knowledge of the cutoff dependence of the QNS to correct for the discretization errors in the pressure. 
In the weak-coupling picture, the pressure can be written as a sum of quark and 
gluon pressures \(p^q\) and \(p^g\), where the latter has only negligible cutoff 
effects for an improved gauge action. 
Then we may assume that the cutoff dependence is completely carried by \(p^q\), 
which is dominated by the light quarks, and that it ought to be similar to the 
cutoff dependence of the second order light QNS \(\chi^l_2\). 
Finally, we may assume that the continuum limit of \(p^q\) at high temperatures 
can be estimated using the rescaled ideal quark gas pressure \(p^q_{\rm id}\).  
On the grounds of these considerations, we may develop a scheme for removing the discretization errors of the pressure by using the known results for QNS,
\al{
&
p(T)=p(T,N_\tau)+p^q(T) \left(1-\frac{p^q(T,N_\tau)}{p^q(T)}\right),
&&
\frac{p^q(T,N_\tau)}{p^q(T)} \simeq \frac{\chi^l_2(T,N_\tau)}{\chi^l_2(T)},
\quad
p^q(T) \approx 0.85p^q_{\rm id}(T).
}
The advantage of determining corrections from using QNS is that these calculations 
do not rely on the availability of ensembles at zero temperature with the same 
parameters. 
As such, it is considerably cheaper to determine the discretization errors of 
\(\chi^q_{2n}\) than for the pressure. 

\begin{figure*}
\includegraphics[width=7.5cm]{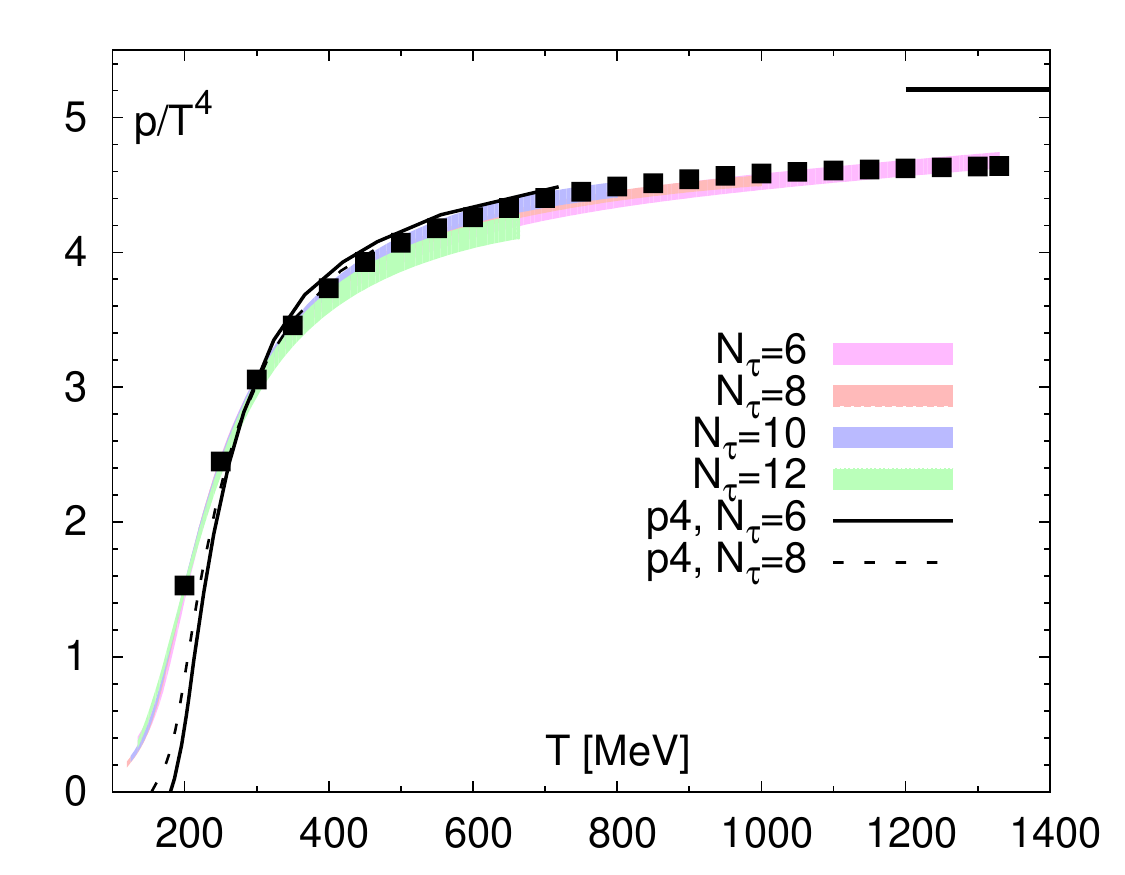}
\includegraphics[width=7.5cm]{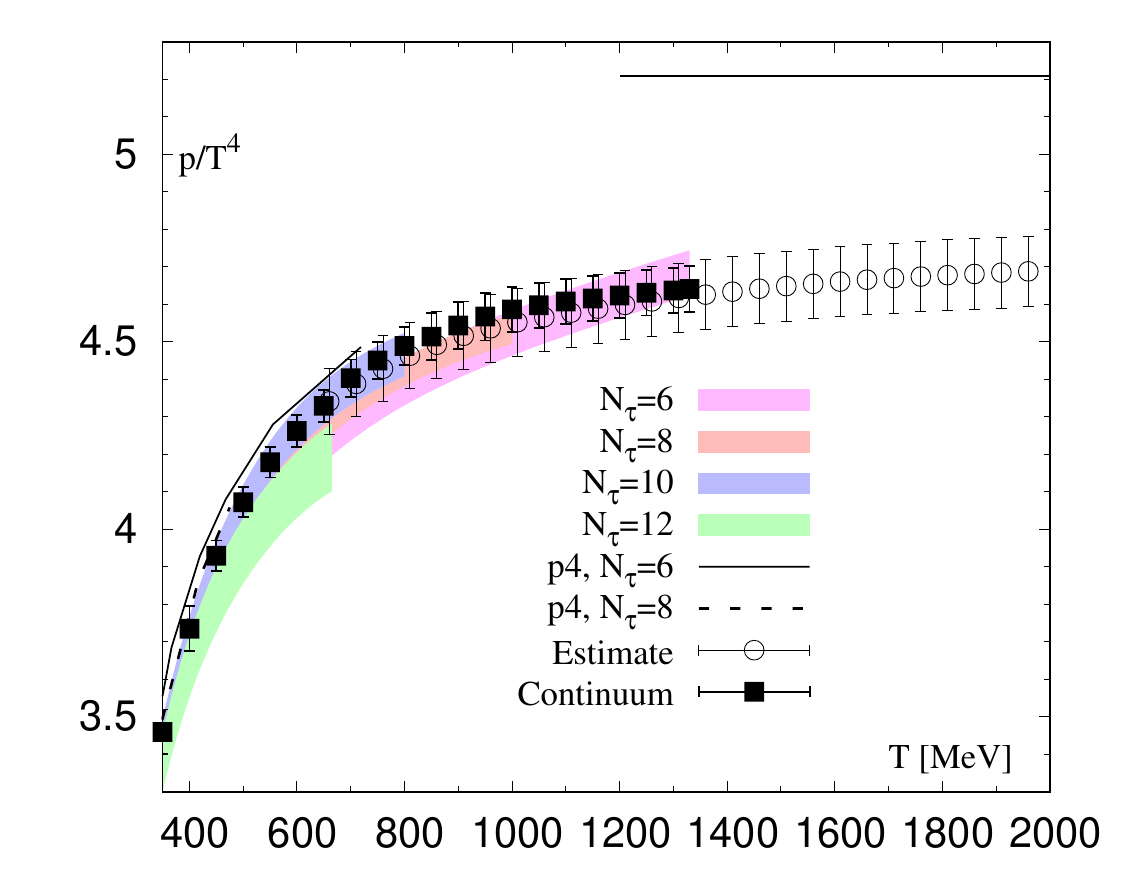}
\caption{
Left:
The pressure calculated with HISQ action for different $N_{\tau}$ and 
corrected for cutoff effects. 
The filled squares are the continuum limit of the pressure.
For comparison we also plot corrected results with p4 action.
Right:
The high temperature continuum estimate (open circles) is consistent with the continuum limit and the corrected pressure even without the conservatively 
enlarged errors. 
}
\label{fig:pcont}
\end{figure*}

We use the second and third approaches to determine the continuum limit and 
results that are corrected for cutoff effects up to \(T = 1.33\,{\rm GeV}\). 
In \mbox{Fig.}~\ref{fig:pcont} we show these results together with results 
obtained with p4 action~\cite{Cheng:2007jq} that have been subjected to the 
same correction approach using corresponding QNS results with p4 action~\cite{Cheng:2008zh}. 
We interpret this as a validation of the correction procedure that the corrected 
results with different actions approach the same continuum limit from opposite sides, 
\mbox{cf. Fig.}~\ref{fig:p_chi}, and are consistent for \(T \gtrsim 400\,{\rm MeV}\) 
with \(N_\tau \geq 8\). 
We note that our continuum result for \(T=500\,{\rm MeV}\) is one and a half 
standard errors larger than the 2+1 flavor result for stout 
action~\cite{Borsanyi:2013bia} and  agrees well with the previous result for HISQ 
action~\cite{Bazavov:2014pvz}, although our result has significantly smaller errors. 
Eventually this is still not sufficient for reaching the highest temperatures 
\(T \approx 2\,{\rm GeV}\) and we need a fourth approach.  
Hence, we use the first approach to obtain a continuum estimate for \(T > 1.33\,{\rm GeV}\). 
Recalling that discretization errors in the trace anomaly with \(N_\tau \geq 8\) 
are small for \(T > 300\,{\rm MeV}\), we obtain a continuum estimate for 
\(\Theta^{\mu\mu}\) by performing a spline interpolation of these data for 
\(T \geq 300\,{\rm MeV}\). 
Comparison of this continuum estimate and the data with \(N_\tau = 6\) or 
\(N_\tau=4\) reveals that these are in the temperature window 
\(800\,{\rm MeV} < T < 1\,{\rm GeV}\) smaller by a factor \(1.4\) or \(1.2\) 
respectively. 
Assuming a mild temperature dependence for the trace anomaly at high temperatures 
as suggested by the weak-coupling limit, 
we use the same rescaling factors \(1.4\) or \(1.2\) respectively to estimate 
corrected results for the trace anomaly from the data with \(N_\tau = 6\) or 
\(N_\tau=4\). 
We then perform a combined spline interpolation of the trace anomaly with \(N_\tau \geq 8\) for \(T \geq 400\,{\rm MeV}\) and the corrected data with \(N_\tau = 6\) or 
\(N_\tau=4\) for \(T>1\,{\rm GeV}\), where we assign a 40\% or 20\% systematic error 
to these corrected data to be conservative. 
Finally we apply the integral method to this continuum estimate and fix the 
reference point through the continuum limit of the pressure at \(T=660\,{\rm MeV}\). 
In order to be sufficiently conservative we double the errors of this continuum 
estimate for the pressure and show it in the right panel of 
\mbox{Fig.}~\ref{fig:pcont}. 
We stress that none of these error enlargements are necessary in order to obtain full 
consistency with the continuum limit or the QNS corrected lattice data over the whole 
common temperature range. 

\section{Comparison to weak coupling}\label{sec:weak}

\begin{figure*}
\includegraphics[width=7.5cm]{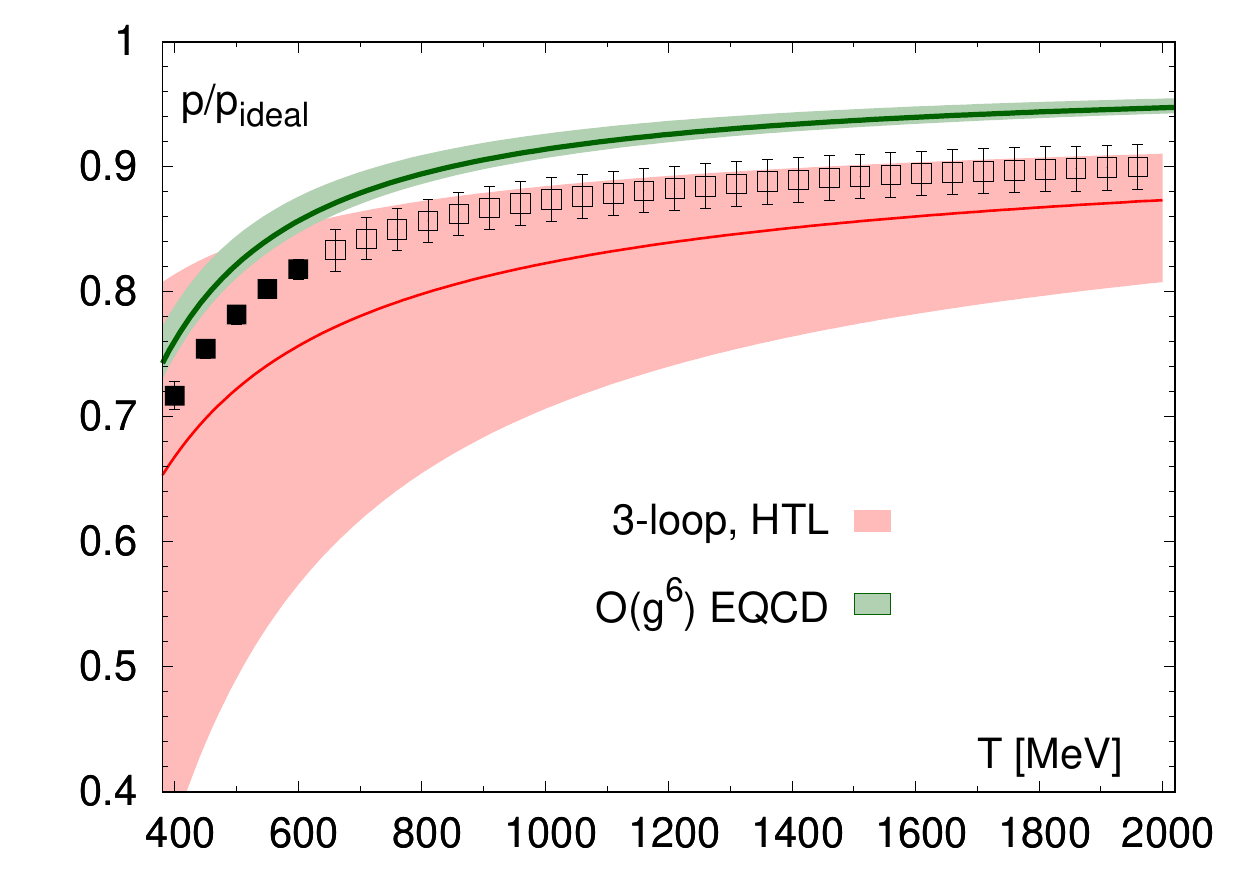}
\includegraphics[width=7.5cm]{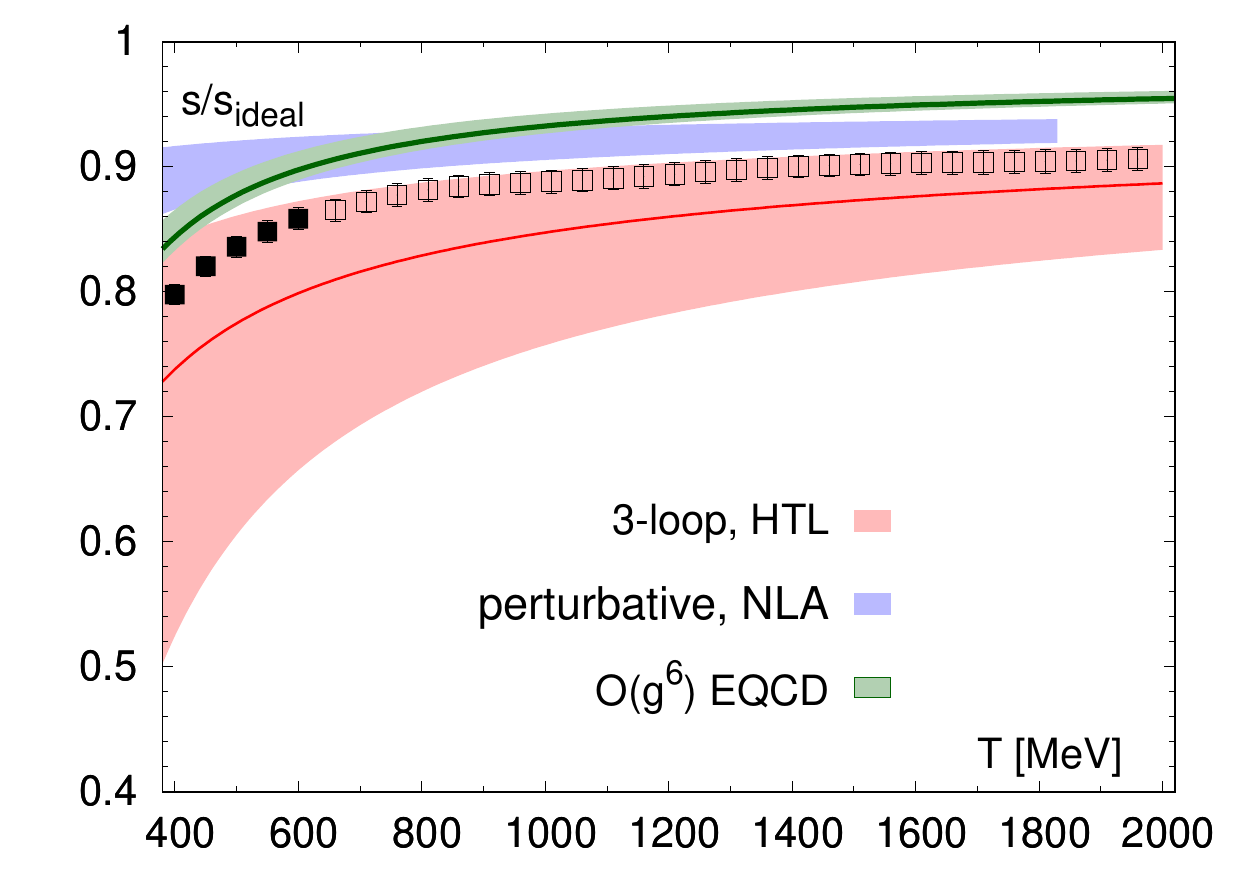}
\caption{The pressure (left) and the entropy density (right) at high
temperatures compared to weak-coupling results.
The filled or open symbols are the continuum limit 
(second approach in \mbox{Sec.}~\ref{sec:pressure}), 
or the continuum estimate 
(fourth approach in \mbox{Sec.}~\ref{sec:pressure}).
We show the HTL~\cite{Haque:2014rua} (red) and EQCD~\cite{Laine:2006cp} (green) results each with a line (\(\mu=2\pi T\)) and a band ($\pi T$ to $4\pi T$).
For the entropy density the blue band corresponds to the resummed calculation in 
next-to-leading log approximation (NLA)~\cite{Rebhan:2003fj}.
}
\label{fig:eos_high}
\end{figure*}

Armed with the continuum result developed throughout the last section we embark on a comparison to the weak-coupling calculations. 
\begin{figure}\center
\includegraphics[width=7.5cm]{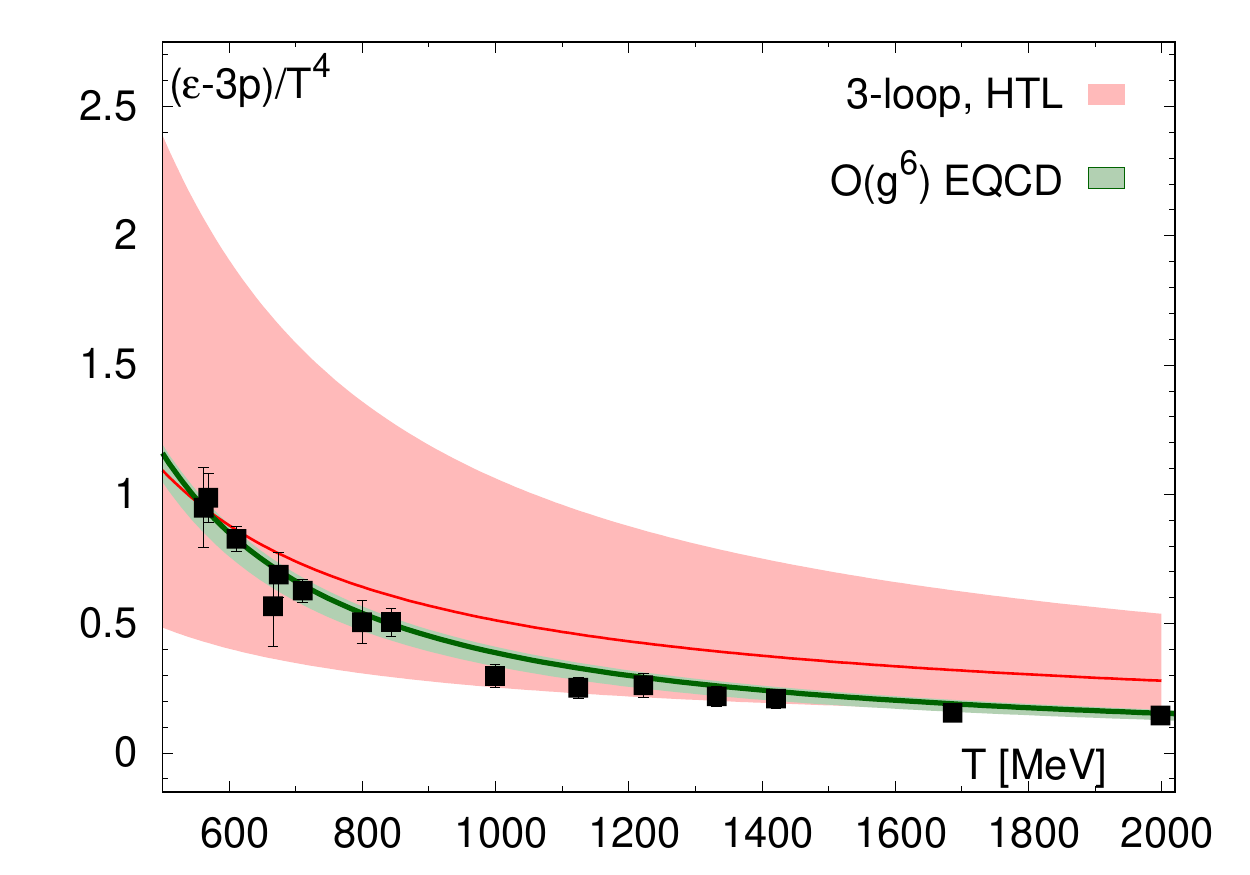}
\caption{
The comparison of the lattice data for the trace anomaly with
HTL perturbation theory (red) and EQCD (green) each shown with a line (\(\mu=2\pi T\)) and a band ($\pi T$ to $4\pi T$). 
Data with $N_{\tau}=6$ and $4$ for $T>1\,{\rm GeV}$ have been 
rescaled by $1.4$ and $1.2$, respectively.}
\label{fig:e-3p_comp}
\end{figure}
In \mbox{Fig.}~\ref{fig:e-3p_comp} we show the lattice data for the trace anomaly, 
where data with \(N_\tau=6\) or \(4\) have been corrected, see the discussion in \mbox{Sec.}~\ref{sec:pressure}. 
We show the data together with analytic results obtained using hard thermal loop (HTL) 
perturbation theory at three loop~\cite{Haque:2014rua} and dimensionally reduced 
effective field theory, namely the electrostatic QCD (EQCD) at order 
\(\mathcal{O}(g^6)\)~\cite{Laine:2006cp}.
The uncertainty band of the weak-coupling results are due to variation of the 
resummation scale \(\mu\).  
We see fair agreement between the different calculations. 
In \mbox{Fig.}~\ref{fig:eos_high}, we compare the lattice result for the pressure 
and entropy density \(s=\partial p/\partial T\) to the weak-coupling calculations.
Our result is higher than the central value of the HTL result by about one 
sigma, but still covered by its scale uncertainty, and a few percent lower than the 
EQCD prediction.
We also compare the entropy to the resummed calculation in next-to-leading 
log approximation (NLA)~\cite{Rebhan:2003fj}, which is higher than our result, but 
consistent within uncertainties for \(T>1.3\,{\rm GeV}\). 
From our comparison it is evident that the presented lattice data are sufficiently 
precise in order to determine through a comparison with weak-coupling results, 
whether the thermodynamics of the quark-gluon plasma can be understood in terms of 
the weak-coupling picture in the given temperature range. 

\section{Conclusions}\label{sec:conclusions}

We have extended the calculation of the 2+1 flavor QCD Equation of State with 
the HISQ action to higher temperatures. 
In order to extend the previous result to higher temperatures we have used gauge 
ensembles with a light sea quark mass \(m_l=m_s/5\) and demonstrated that the quark 
mass dependence of the trace anomaly is covered by the statistical uncertainty for 
\(T > 400\,{\rm MeV}\). 
We have determined the pressure from the trace anomaly using the integral method. 
At low temperatures, we have used a hadron resonance gas with distorted spectrum as 
reference and have demonstrated that it can describe the trace anomaly for 
\(T<140\,{\rm MeV}\). 
We have studied the cutoff dependence of the pressure at low and high temperatures 
and extrapolated the pressure to the continuum limit. 
Moreover, we have developed a correction scheme for the cutoff effects in the 
pressure based on the known cutoff dependence of the quark number susceptibilities 
and the known weak-coupling limit at high temperatures. 
We demonstrated the validity of this correction scheme by applying it to results 
with the HISQ and p4 actions at finite \(N_\tau\), which become numerically 
consistent with the continuum limit after correction. 
Finally, we have developed another correction scheme for the trace anomaly with 
\(N_\tau<8\) at high temperatures, have demonstrated that it agrees with the other 
results and have extracted a continuum estimate up to \(T \approx 2\,{\rm GeV}\). 
We have compared this estimate including conservative errors to weak-coupling 
predictions for the trace anomaly, the pressure and the entropy density and 
find fair agreement at high temperatures. 
The lattice results for pressure and entropy density are in the middle between the 
predictions of different weak-coupling calculations. 

\section*{Acknowledgments}

This research was supported by the DFG cluster of excellence "Origin and Structure 
of theUniverse" (\href{www.universe-cluster.de}{www.universe-cluster.de}).
The simulations have been carried out on the computing facilities of the 
Computational Center for Particle and Astrophysics (C2PAP), SuperMUC and NERSC. 
We used the publicly available MILC code to perform the numerical
simulations~\cite{milc}. 
The data analysis was performed using the R statistical package~\cite{Rpackage}. 
This work has been supported in part by the U.S. Department of Energy 
through grant Contract No. DE-SC0012704. 
J. H. W. acknowldges the support by the Bundesministerium f\"ur Bildung 
und Forschung (BMBF) under Grant No. ``Verbundprojekt 05P2015--ALICE at High 
Rate (BMBF-FSP 202) GEM-TPC Upgrade and Field theory based investigations 
of ALICE physics'' under Grant No. 05P15WOCA1.
J. H. W. would like to thank Y. Schr\"oder for helpful discussions and for providing 
updated EQCD results.


\begin{thebibliography}{99}

\bibitem{Cheng:2007jq} 
  M.~Cheng {\it et al.},
  Phys.\ Rev.\ D {\bf 77}, 014511 (2008)
  [arXiv:0710.0354 [hep-lat]].

\bibitem{Bazavov:2009zn} 
  A.~Bazavov {\it et al.},
  Phys.\ Rev.\ D {\bf 80}, 014504 (2009)
  [arXiv:0903.4379 [hep-lat]].
  
\bibitem{Borsanyi:2010cj} 
  S.~Borsanyi, G.~Endrodi, Z.~Fodor, A.~Jakovac, S.~D.~Katz, S.~Krieg, C.~Ratti and K.~K.~Szabo,
  JHEP {\bf 1011}, 077 (2010)
  [arXiv:1007.2580 [hep-lat]].

\bibitem{Borsanyi:2013bia} 
  S.~Borsanyi, Z.~Fodor, C.~Hoelbling, S.~D.~Katz, S.~Krieg and K.~K.~Szabo,
  Phys.\ Lett.\ B {\bf 730}, 99 (2014)
  [arXiv:1309.5258 [hep-lat]].

\bibitem{Bazavov:2014pvz} 
  A.~Bazavov {\it et al.} [HotQCD Collaboration],
  Phys.\ Rev.\ D {\bf 90}, 094503 (2014)
  [arXiv:1407.6387 [hep-lat]].

\bibitem{Borsanyi:2016ksw} 
  S.~Borsanyi {\it et al.},
  Nature {\bf 539}, no. 7627, 69 (2016)
  [arXiv:1606.07494 [hep-lat]].
  
\bibitem{Bazavov:2017dsy} 
  A.~Bazavov, P.~Petreczky and J.~H.~Weber,
  Phys.\ Rev.\ D {\bf 97}, no. 1, 014510 (2018)
  [arXiv:1710.05024 [hep-lat]].

\bibitem{Follana:2006rc} 
  E.~Follana {\it et al.} [HPQCD and UKQCD Collaborations],
  Phys.\ Rev.\ D {\bf 75}, 054502 (2007)
  [hep-lat/0610092].
  
\bibitem{Bazavov:2013pra} 
  A.~Bazavov {\it et al.} [MILC Collaboration],
  PoS LATTICE {\bf 2013}, 154 (2014)
  [arXiv:1312.5011 [hep-lat]].

\bibitem{Laine:2006cp} 
  M.~Laine and Y.~Schroder,
  Phys.\ Rev.\ D {\bf 73}, 085009 (2006)
  [hep-ph/0603048].
  
\bibitem{Haque:2014rua} 
  N.~Haque, A.~Bandyopadhyay, J.~O.~Andersen, M.~G.~Mustafa, M.~Strickland and N.~Su,
  JHEP {\bf 1405}, 027 (2014)
  [arXiv:1402.6907 [hep-ph]].

\bibitem{PPJHW:as2018}
  P.~Petreczky and J.~H.~Weber,

\bibitem{Bazavov:2011nk} 
  A.~Bazavov {\it et al.},
  Phys.\ Rev.\ D {\bf 85}, 054503 (2012)
  [arXiv:1111.1710 [hep-lat]].
  
\bibitem{Bazavov:2016uvm} 
  A.~Bazavov, N.~Brambilla, H.-T.~Ding, P.~Petreczky, H.-P.~Schadler, A.~Vairo and J.~H.~Weber,
  Phys.\ Rev.\ D {\bf 93}, no. 11, 114502 (2016)
  [arXiv:1603.06637 [hep-lat]].
  
\bibitem{Bazavov:2018wmo} 
  A.~Bazavov {\it et al.} [TUMQCD Collaboration],
  Phys.\ Rev.\ D {\bf 98}, no. 5, 054511 (2018)
  [arXiv:1804.10600 [hep-lat]].

\bibitem{Bazavov:2013uja} 
  A.~Bazavov {\it et al.},
  Phys.\ Rev.\ D {\bf 88}, no. 9, 094021 (2013)
  [arXiv:1309.2317 [hep-lat]].

\bibitem{Cheng:2008zh} 
  M.~Cheng {\it et al.},
  Phys.\ Rev.\ D {\bf 79}, 074505 (2009)
  [arXiv:0811.1006 [hep-lat]].

\bibitem{Rebhan:2003fj} 
  A.~Rebhan,
  hep-ph/0301130.

\bibitem{milc}
 MILC collaboration, 
 MILC code, 
 \href{http://www.physics.utah.edu/~detar/milc/}{http://www.physics.utah.edu/$\sim$detar/milc/}.
    
\bibitem{Rpackage} 
  {R Core Team},
  {R: A Language and Environment for Statistical Computing},
  \href{https://www.R-project.org}{https://www.R-project.org}.

\end{thebibliography}
\end{document}